\documentclass[aps,twocolumn,groupedaddress,floatfix]{revtex4-1}
\usepackage{amssymb,amsmath}

\DeclareMathOperator{\arcsinh}{arcsinh}

\DeclareMathOperator{\arccoth}{arccoth} 
\usepackage{graphicx}
\usepackage{subfigure}
\usepackage[english]{babel}
\usepackage{float}
\usepackage{color}
\usepackage[document]{ragged2e}

\usepackage{lipsum}
\begin{document}
\newcommand{\be}{\begin{equation}}
\newcommand{\ee}{\end{equation}}
\newcommand{\rojo}[1]{\textcolor{red}{#1}}

\title{Interplay of fractionality and $\cal{PT}$- symmetry on a 1D lattice}

\author{Mario I. Molina}
\affiliation{Departamento de F\'{\i}sica, Facultad de Ciencias, Universidad de Chile, Casilla 653, Santiago, Chile}

\date{\today }

\begin{abstract} 
We examine the stability domains of a 1D discrete Schr\"{o}dinger equation in the simultaneous presence of 
parity-time ($\cal{PT}$) symmetry and fractionality. Direct numerical examination of the eigenvalues of the system reveals that, as the fractional exponent is decreased away from unity (the standard case), the instability gain increases abruptly past a critical value. Also, as the length of the system  increases, the stable fraction decreases as well. Also, for a fixed fractional exponent and lattice size, an increase in gain/loss also brings about an abrupt increase in the instability gain. Finally, the participation ratio of the modes is seen to decrease with an increase of the gain/loss parameter and with a decrease of the fractional exponent, evidencing a tendency towards localization.

\end{abstract}

\maketitle

{\em Introduction}. Two physics developments have called for increased attention in recent times. One is the phenomenon of $\cal{P}{\cal T}$ symmetry, and the other is fractionality. Parity-time ($\cal{P}{\cal T}$) systems are characterized for having a non-hermitian Hamitonian, but a real spectrum nonetheless. This happens for a Hamiltonian that is invariant with respect to the simultaneous action of 
parity inversion and time reversal. Typically, the spectrum remains real until the gain/loss parameter surpasses a critical value. At that point a pair of  eigenvalues become complex rendering the dynamics unstable\cite{bender1,bender2}. It is said then that the ${\cal P}{\cal T}$ symmetry is then spontaneously broken\cite{optics4}.

The field of $\cal{P}{\cal T}$ symmetry quickly found a realization in optics, where for onedimensional systems it was ascertained that for the system to obey ${\cal P}{\cal T}$, the imaginary (real) part of the index of refraction needed to be an odd (even) function in space. Under these conditions a balanced gain and loss is possible. Currently, numerous $\cal{P}\cal{T}$-symmetric systems have been explored in several settings, from electronic circuits\cite{circuits}, optics\cite{optics1,optics2,optics3,optics4,optics5}, magnetic metamaterials\cite{MM}, to solid-state and atomic physics\cite{solid1,solid2}, among others. The ${\cal P}{\cal T}$ symmetry-breaking phenomenon has been observed in several experiments\cite{optics5,experiment2,experiment3}.

The second recent development is that of fractional physics which extends the usual integer calculus to a fractional one, with its definitions of a fractional integral and fractional derivative.
This topic has a long history dating back to the observation that the derivative $d^{n} x^{k}/d x^{n} = k!/(k-n)!\ x^{k-n}$ for integer $n$ could be extended to non-integer orders by means of the Gamma function: $d^{\alpha} x^{k}/d x^{\alpha} = \Gamma(k+1)/\Gamma(k-\alpha+1)\ x^{k-\alpha}$. From that point, rigorous work done by several people, including Riemann,  Euler, Laplace, Caputo and others, have transformed fractional calculus from a mathematical curiosity into a serious research field. Several possible definitions for the fractional derivative have been advanced, each one with its own advantages and disadvantages. One of the most used definitions is the Riemann-Liouville form
\be
\left( {d^{\alpha}\over{dx^{\alpha}}} \right) f(x) = {1\over{\Gamma(1-\alpha)}} {d\over{dx}} \int_{0}^{x} {f(s)\over{(x-s)^\alpha}},
\ee
where $0<\alpha<1$. The non-local character of the fractional derivative has proven useful in a variety of fields: fluid mechanics\cite{quantum}, fractional kinetics and anomalous diffusion\cite{kinetics1,kinetics2,kinetics3}, strange kinetics\cite{strange}, fractional quantum mechanics\cite{frac1,frac2}, Levy processes in quantum mechanics\cite{levy}, plasmas\cite{plasmas}, electrical propagation in cardiac tissue\cite{cardiac}, biological invasions\cite{invasions}, and epidemics\cite{epidemics}.

In this work we examine the interplay between ${\cal P}{\cal T}$ and fractionality. In particular, it is interesting to ascertain the stability regions in gain/loss and fractional exponent space. As we will see,  as the fractional exponent is decreased away from unity (the standard case), the instability gain increases abruptly past a critical value, i.e., we enter an unstable phase with the presence of complex eigenvalues. Something similar happens when the fractional exponent is kept fixed and the gain/loss coefficient is increased: The instability gain abruptly increases past a certain value. Also, and in agreement with old computations of ${\cal P}{\cal T}$ for a 1D chain we observe a quick decrease of the stability region with an increase in system length. 

{\em The model}.\ \ We start from a 1D tight-binding model that contains ${\cal P}{\cal T}$ symmetry:
\be
 i {d C_{n}\over{d t}} + V(C_{n+1} + C_{n-1}) + \epsilon_{n} C_{n} = 0\label{1}
 \ee
 where $\epsilon_{n}$ is a complex quantity whose imaginary (real) part is odd (even) in space. For example,
 \be
 \epsilon_{n}=
\begin{cases}
\hphantom{-}i \gamma &,\, \text{if n\ odd},\\
-i \gamma&,\, \text{if n\ even.}
\end{cases}\label{2}
\ee
From here on, we will take the real part of $\epsilon_{n}$ as zero. The parameter $\gamma$ is called the gain/loss coefficient and determines the balance between gains and losses in the system. For systems such as (\ref{1}),(\ref{2})
it was shown a long time ago that, in the limit of an infinite chain, the system is always 
in the broken phase, i.e., unstable (complex eigenvalues)\cite{MM,broken2}. However, it has been shown that for finite arrays, a region of stability (real eigenvalues) could be  possible\cite{stable}.  Configuration (\ref{2}) corresponds to the sequence $\cdots,+,-,+,-,+,-,+,-,+,-,+,-\cdots$. But other types of simple gain/loss distributions are possible.
For instante, $\cdots,+, -,+, -, +, 0, -, +, -, +, -,\cdots$, or even the distribution $\cdots, -,-,-,-,-,-,0,+,+,+,+,+,+,\cdots$. For this last case, we will
see that, even though the concentration of loss and gain values on
opposite sides, the dynamics does possess a stability window for finite arrays lengths.
As we will show, all of these gain/loss distributions lead to similar stability behaviors.
\begin{figure}[t]
 \includegraphics[scale=0.3]{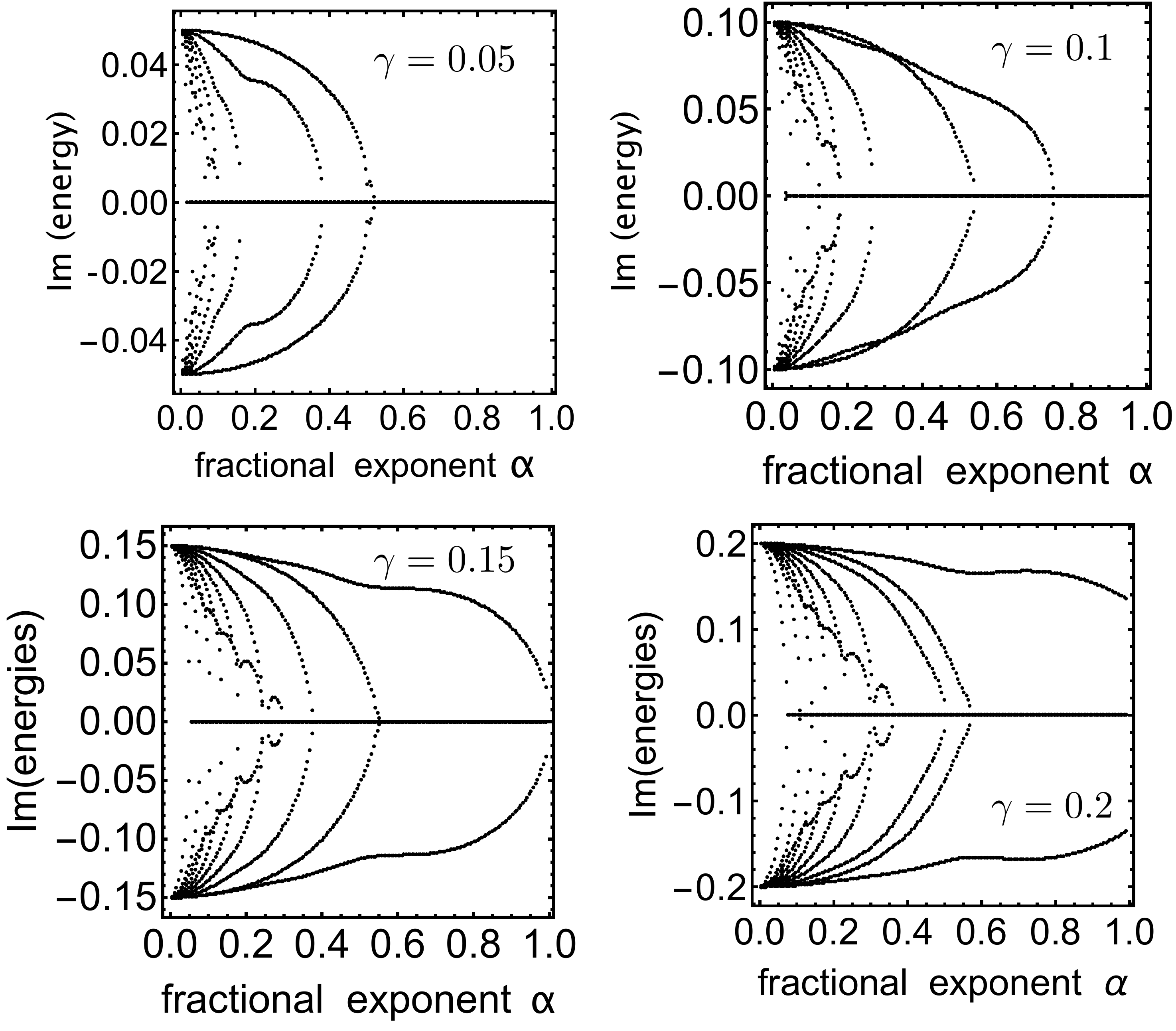}
  \caption{Imaginary part of the eigenvalues as a function of the fractional exponent $s$, for several increasing values of the gain/loss parameter $\gamma$. The gain/loss distribution is $(-1)^n \gamma$ and $N=20$.}
  \label{fig1}
\end{figure}
Now let us go back to main Eq.(\ref{1}). The kinetic energy term $V(C_{n+1}+ C_{n-1})$, is essentially a discrete Laplacian $\Delta_{n} = C_{n+1}-2 C_{n}+C_{n-1}$, so that Eq(\ref{1})can be cast as
\be
 i {d C_{n}\over{d t}} + 2 V C_{n}+ V \Delta_{n} C_{n} + \epsilon_{n} C_{n} = 0\label{4}
\ee
We now proceed to replace the discrete Laplacian $\Delta_{n}$ by its fractional form $(\Delta_{n})^\alpha$ in Eq. (\ref{4}). The closed-form of this fractional discrete Laplacian is given in closed form by\cite{ciaurri}
\be
(-\Delta_{n})^\alpha C_{n}= \sum_{m\neq n} K^\alpha
(n-m) (C_{n}-C_{m}), \hspace{0.5cm} 0<\alpha<1\label{5}
\ee
where,
\be K^{\alpha}(m)={{4^\alpha \Gamma (\alpha+(1/2))\over{\sqrt{\pi}|\Gamma(-\alpha)|}}}{\Gamma(|m|-\alpha)\over{\Gamma(|m|+1+\alpha)}},\label{6}
\ee
and $\Gamma(x)$ is the Gamma function and $\alpha$ is the fractional exponent. We see that the presence of fractionality introduces nonlocal interactions via the symmetric kernel $K^\alpha (n-m)$. After replacing (\ref{5}) into (\ref{4}), and after looking for stationary-state modes  $C_{n}(t) = \phi_{n} \exp(i \lambda t)$, we obtain a system of coupled difference equations for the $\{\phi_{n}\}$
\be
(-\lambda + 2 V + \epsilon_{n} ) \phi_{n} + V \sum_{m\neq n}K^\alpha
(n-m)(\phi_{m}-\phi_{n})=0.\label{9}
\ee
For $n=1$ and $n=N$ we must replace the $2 V$ term in Eq. (\ref{9}) by $V$. 
The long-distance asymptotic behavior can be obtained from Eq.(\ref{6}) and the relations
$\Gamma(n+\alpha) \approx \Gamma (n)\ n^\alpha$ valid at large $n$.  From this we obtain the asymptotic behavior $K^\alpha (m) \approx 1/|m|^{1+2 \alpha} $i.e., an algebraic decay.  Thus, the effective coupling goes as $1/|m|^3$ in the standard case ($\alpha=1$), which is reminiscent of a dipole-dipole interaction, while in the opposite case ($\alpha\sim 0$), the coupling decreases extremely slow as $1/|m|$, meaning that all sites become essentially coupled. 

In the absence of gain/loss $\epsilon_{n}=0$, the dispersion relation can be obtained in closed form by inserting a plane wave solution $\phi_{n}=A\ \exp(i k n)$ into Eq.(\ref{9}), obtaining

\begin{widetext}
\be
\lambda(k) = 2V - 4 V \sum_{m=1}^{\infty} K^\alpha(m) \sin((1/2) m k)^2\label{dispersion}
\ee
or, in closed form\cite{dnls}
\be
\lambda(k)=2 V - {16 V\ \Gamma(\alpha+(1/2))\over{\sqrt{\pi}\ \Gamma(1+\alpha)}}\Big( 1-\exp(-i k)\ \alpha\ \Gamma(1+\alpha)[\ R(1,1-\alpha,2+\alpha;\exp(-i k))+\exp(2 i k) \ R(1,1-\alpha,2+\alpha;\exp(i k))\ ] \Big)\label{dispersion2}
\ee
\end{widetext}
where $R(a,b,c;z)={}_2 F _{1}(a,b,c;z)/\Gamma(c)$ is the regularized hypergeometric function.

As was shown in ref.\cite{dnls}, the bandwidth decreases with decreasing $\alpha$ until at $\alpha\rightarrow 0$, the band becomes flat with all modes degenerate. For general gain/loss distributions $\epsilon_{n}$, a numerical solution of Eq.(\ref{9}) must be computed.
\begin{figure}[t]
 \includegraphics[scale=0.5]{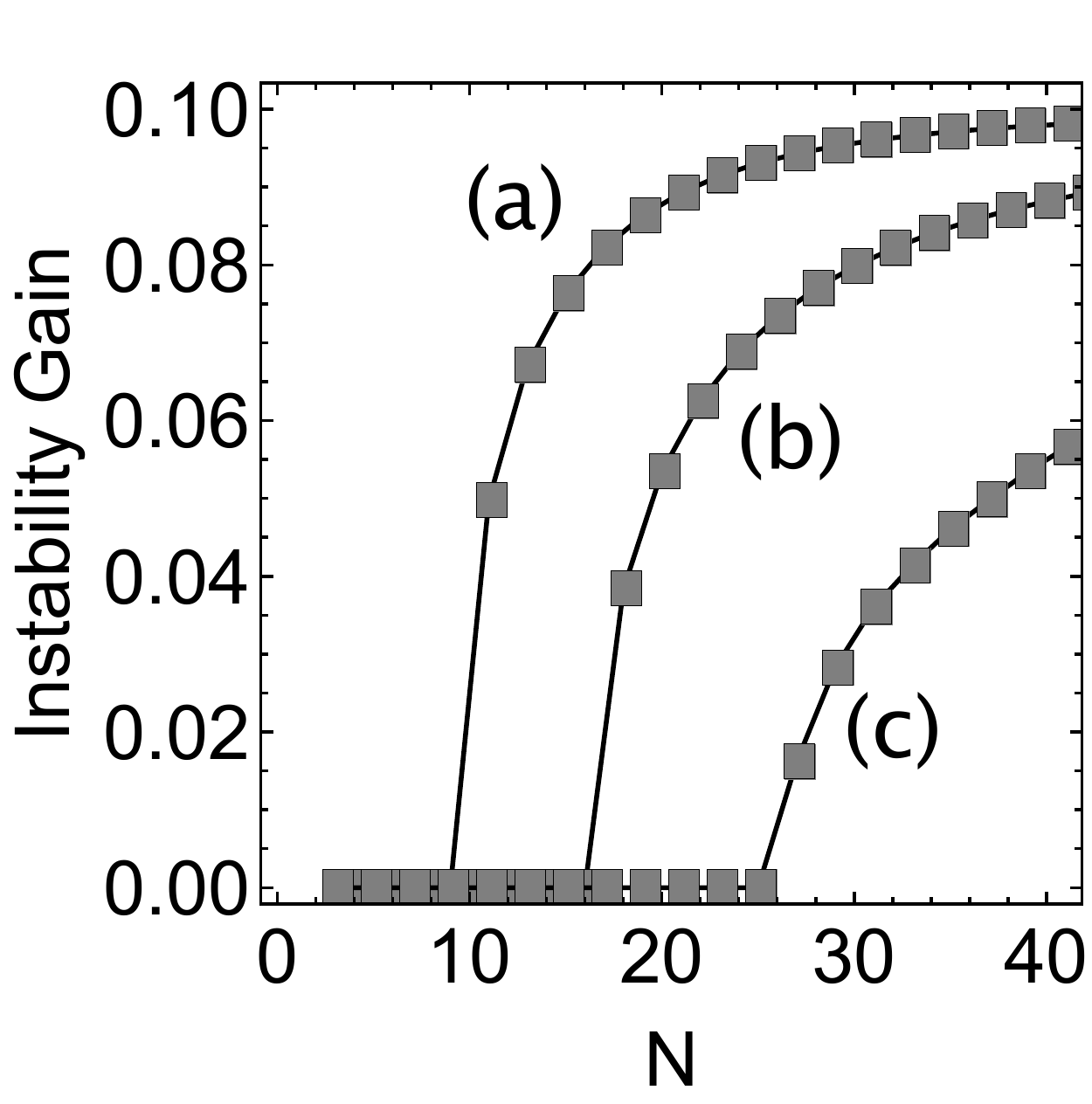}
  \caption{Instability gain $G$ versus $N$, for three different gain/loss distributions and
  (a) $\alpha=0.8,\gamma=0.1$ (b) $\alpha=0.6,\gamma=0.1$ and (c) $\alpha=0.7,\gamma=0.1$. The gain/loss distributions used are (a) $\cdots,-\gamma,-\gamma,-\gamma,0,\gamma,\gamma,\gamma,\gamma\cdots$},\hspace{-2.5cm}(b) $\cdots,\gamma,-\gamma,\gamma,-\gamma,\gamma,-\gamma,\gamma,-\gamma,\cdots$ and \\
(c) $\cdots,\gamma,-\gamma,\gamma,-\gamma,0,\gamma,-\gamma,\gamma,-\gamma,\cdots$
  \label{fig2}
\end{figure}

{\em Results}.\ \ Let us proceed to compute the stability of the lattice under the combined influence of fractionality and ${\cal P}{\cal T}$ symmetry. To this end we fix values of $N$, $\gamma$ and $\alpha$ and compute the eigenvalues of the system. When all eigenvalues are real, the system's dynamics is bounded; however, if at least a couple of (complex conjugate) eigenvalues is complex, an oscillating instability will appear and the dynamics will be unbounded. Figure 1 shows a plot of the imaginary part of all eigenvalues as a function of the fractional parameter $\alpha$, for four fixed values of the gain/loss parameter $\gamma$. As we can see, in all cases there is a fractional exponent range inside which the eigenvalues are purely real, meaning a stable behavior. This range decreases, however, as the gain/loss coupling is augmented and, at a certain finite $\gamma$ value, all the eigenvalues acquire an imaginary part. At this point, the system suffers a {$\cal{PT}$} symmetry-breaking transition going into the unstable regime.

The stability behavior is monitored through the instability gain $G$, defined as 
$G=\mbox{Max}|\{\mbox{Im}(\lambda_{n})\}|$ for a given $N, \alpha$ and $\gamma$. Figure 2 shows an example of this instability gain as a function of system size $N$, for three different fractional and gain/loss distributions. In all cases we appreciate a sudden transition form stability ($G=0$) to instability ($G>0$) as $N$ is increased. This transition to instability occurs sooner for distribution (a). This can be explained as the effect of having the negative sites far from the positive ones, which facilitates the accumulation of energy on the positive sector. For cases (b) and (c), the alternation of positive and negative sites, reduce the possibility of energy accumulation. In any case,  
our 1D system is unstable in the large $N$ limit, which is in agreement with previous work\cite{broken2}. The reason for this instability with $N$ can be understood by the following rough argument: 
\begin{figure}[t]
 \includegraphics[scale=0.4]{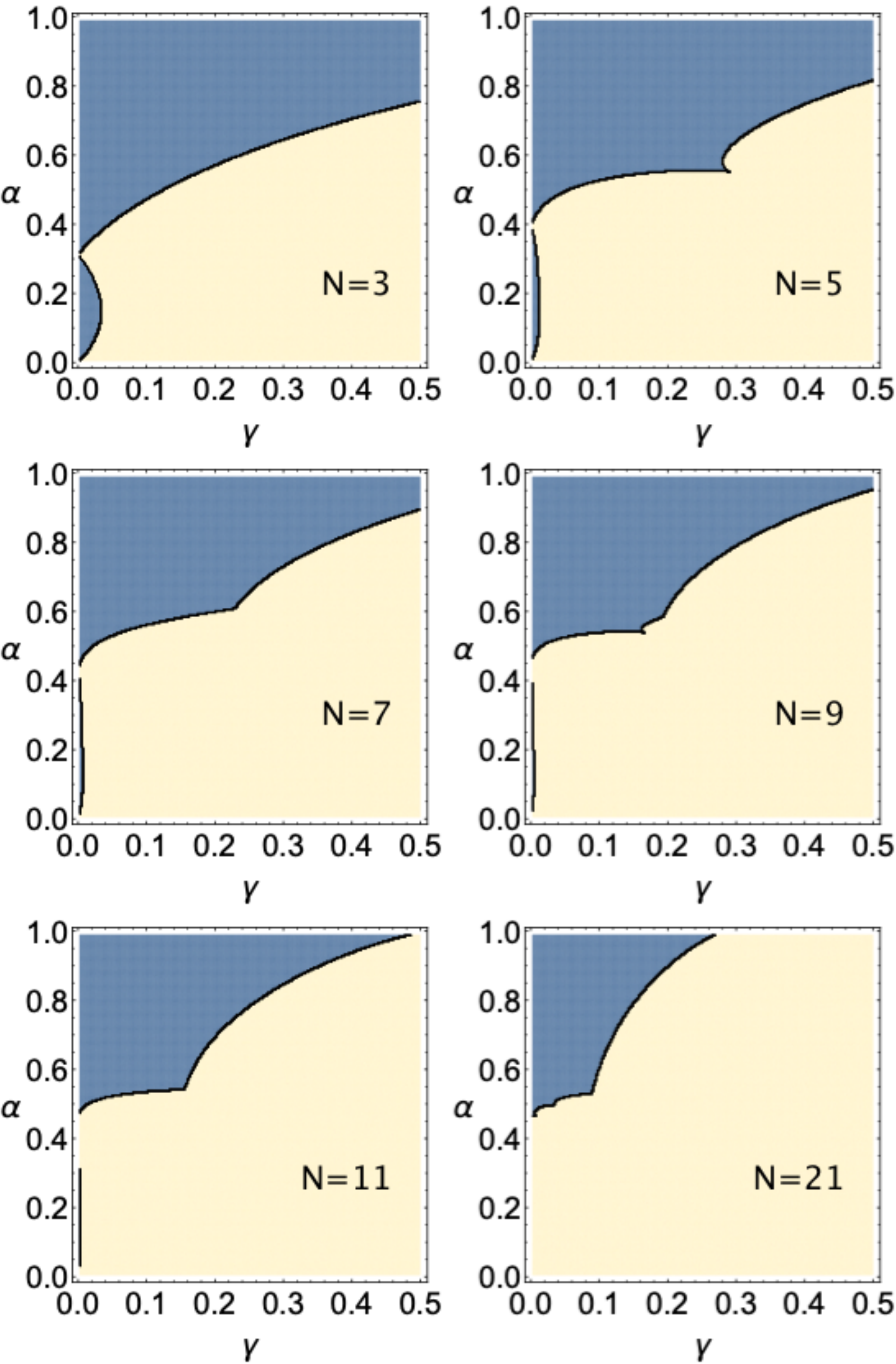}
  \caption{Instability gain $G$ as a function of the fractional exponent $\alpha$ and the gain/loss parameter $\gamma$, for different lattice sizes. The dark (clear) regions denote stable (unstable) regimes.}
  \label{fig3}
\end{figure}
The idea is to compare the time needed for energy to transfer from one site to a neighboring site, compared to the time employed by the site to accumulate energy. For a large array, and in the absence of gain/loss  effects, we have the dispersion relation for the waves: $\Omega_{k}=2 V - 4 V \sum_{n} K^{\alpha}(n) \sin((1/2) n k)^2$. The group velocity of these waves will be
\be
v_{k}= {d \Omega_{K}\over{d k}}=-2 V \sum_{n} n K^{\alpha}(n) \sin(n k).\label{speed}
\ee
Now, for a site with gain, the amplitud grows in time as $\exp(\beta t)$. Therefore, in order for a wave with wavevector 
$k$ to be stable, its velocity $|v_{k}|$ needs to be greater than the speed at which the site accumulates energy: $|v_{k}|>\beta$. In order for the whole system to be stable, one needs this to hold for every $k$. In particular it should hold for the slowest mode, $k\ll 1$. For these modes one has for a periodic array $k=\pi/(N-1)$. This implies
\begin{eqnarray}
\beta& < & {2 \pi V\over{N}}\  \sum_{n=1}^N n^2 K^{\alpha}(n)\nonumber\\
     &   & ={(1+N)\alpha \Gamma(1+N-\alpha)\Gamma(2 \alpha)\sin(\pi \alpha)\over \pi (1-\alpha)\  \Gamma(1+N+\alpha).} 
\end{eqnarray}
From this expression it is possible to show that, at large $N$, $\beta\rightarrow 0 $ for 
$\alpha>1/2$. Thus, the infinite 1D chain will always be in the broken ${\cal P}{\cal T}$ phase for $\alpha>1/2$.  Numerical examination of $\beta$ for $0<\alpha<1/2$, shows that the system is also unstable in this case. These results suggest that, in large versions of our arrays there is no time for the accumulated energy to be transferred away from a `gain' site to neighboring `loss' sites, thus causing the instability. 
 
A bird's-eye view of the system stability as a function of its fractional and gain/loss parameters, is shown in Fig.3. The dark(clear) shaded areas corresponds to stable (unstable) regimes, for several system sizes $N$.  Several behaviors become apparent from this plot: For a fixed gain/loss coefficient $\gamma$, a decrease in fractional exponent $\alpha$ will eventually lead the system into instability. On the other hand, for fixed $\alpha$, an increase in $\gamma$ will also lead the unstable regime eventually. Also, as $N$ is increased, the unstable fraction increases, leading eventually to a completely unstable system for a large (but finite) $N$.

Another interesting observable for our system is the localization behavior of the modes, in the simultaneous presence of $\alpha$ and $\gamma$. A common indicator of localization is the participation ratio $R$ defined as
\be 
R(\alpha,\gamma) = \left\langle {(\sum_{n} |\phi_{n}|^2)^2 \over{\sum_{n} |\phi_{n}|^4}}\right\rangle_{\phi}
\ee
where, for a fixed $\alpha, \gamma$, an average over the $N$ states is taken. For a delocalized state, $R\rightarrow N$,  while for a completely localized one, $R\rightarrow 1$. We take $N=21$ and compute $R$ as a function of the gain/loss parameter $\gamma$, for several fractional exponent values. Results are shown in Fig.4, which shows $R$ for two, different spatial distributions of the gain/loss parameter. For both cases, we see that for a given $\alpha$, the participation ratio $R$ decreases with increasing $\gamma$, with a slope that decreases as $\alpha\rightarrow 0$. For a fixed $\gamma$ value, $R$ decreases sharply with decreasing $\alpha$. This suggests a general tendency of the modes towards localization with increasing (decreasing) gain/loss (fractional exponent).

{\em Discussion}.\ We have examined the interplay of fractionality with ${\cal P}{\cal T}$ symmetry in a simple 1D discrete 
\begin{figure}[t]
 \includegraphics[scale=0.25]{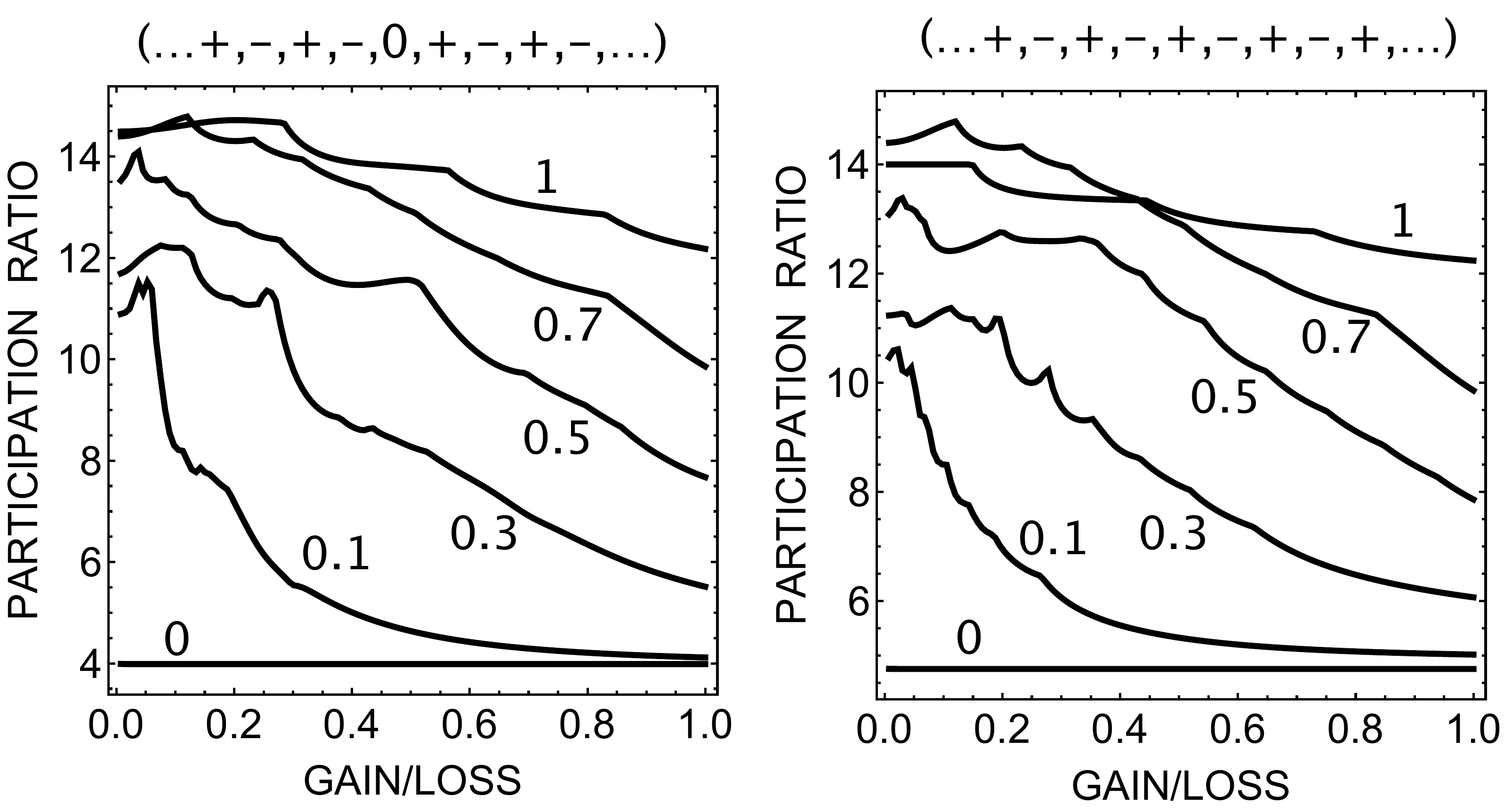}
  \caption{   
Participation ratio $R$ averaged over all modes, as a function of gain/loss parameter. The number on each curve denotes the value of the fractional exponent. On top of each plot we show the gain/loss spatial distribution used. ($N=21$).
  \label{fig4}}
\end{figure}
tight-binding model. By means of a direct numerical computation we have calculated the eigenvalues of the system, and defined an instability gain that characterizes its stability behavior. In general, we find that both, fractionality and gain/loss effects tend to lead the system into the unstable phase. For a fixed gain/loss value, a decrease in fractional exponent from its standard value causes an abrupt transition to instability at certain value. On the other hand, for a fixed fractional exponent, an  increase in gain/loss coefficient also causes an abrupt transition to instability at a given value. Finally, for given $\alpha, \gamma$ values, an increase of the lattice size $N$ also 
leads to an abrupt transition at certain $N$ value. Thus, the infinite 1D chain is always unstable, in agreement with previous related work\cite{broken2}. An examination of the average participation ratio shows a general tendency towards localization with both, a decrease of the fractional exponent and an increase in gain/loss.

We advanced a rough argument that explains the main features of this phenomenon. It is based on the idea that the presence of instability is connected to the inability of the `gain'  sites to transfer their excess energy to `loss' sites quickly enough. A result of this analysis shows that for $\alpha>1/2$, the system will be unstable in the large $N$ limit. For smaller exponent values, $0<\alpha<1/2$, numerical computations shows that the same unstable behavior occurs.

We are currently designing an extension of this work to 2D, where we expect that the system will be even more unstable than in 1D. This is based on the observation that the average distance between two points on a lattice,
\be
D(N,d) = {1\over{N^{2 d}}}\sum_{{\bf n},{\bf m}} |{\bf n} - {\bf m}|
\ee
is greater in 2D than in 1D: 
\begin{eqnarray}
D(N,2)&=&{N\over{120}}(8(2+\sqrt{2})-5\arccoth(\sqrt{2})+45 \arcsinh(1))\nonumber\\
      & & \approx 0.52\ N, 
\end{eqnarray}
compared to $D(N,1)= (1/3)\ N$. The coupling contains the factor $V$ that decreases with the distance between sites. Thus, the average coupling between arbitrary points is smaller in 2D than in 1D. This implies a smaller rate of transfer between points which facilitates the accumulation of energy on a site.

\acknowledgments
This work was supported by Fondecyt Grant 1200120.

\end{document}